\documentclass[prd,twocolumn,nofootinbib,aps,tightenlines,preprintnumbers, superscriptaddress]{revtex4}

\usepackage{epsfig,latexsym,cancel,amssymb,amsmath,verbatim,mathrsfs}
\usepackage{color}
\usepackage{graphicx}

\newcommand{\be}{\begin{equation}}
\newcommand{\ee}{\end{equation}}
\newcommand{\bea}{\begin{eqnarray}}
\newcommand{\eea}{\end{eqnarray}}
\newcommand{\ba}{\begin{array}}
\newcommand{\ea}{\end{array}}

\newcommand{\cpv}{{\textrm{\fontsize{4}{11}\selectfont CP}\!\!\!\!\!\!\diagup}}

\long\def\symbolfootnote[#1]#2{\begingroup%
\def\thefootnote{\fnsymbol{footnote}}\footnote[#1]{#2}\endgroup}

\newcommand{\beq}{\begin{equation}}
\newcommand{\eeq}{\end{equation}}

\newcommand{\tev}{\, {\rm TeV}}

\begin{document}


\title{Anomalous B meson mixing and baryogenesis\\ in a two Higgs doublet model with top-charm flavor violation} %

\author{Sean Tulin} 
\affiliation{Theory Group, TRIUMF, 4004 Wesbrook Mall, Vancouver, BC, 
V6T 2A3, Canada}
\author{Peter Winslow} 
\affiliation{Theory Group, TRIUMF, 4004 Wesbrook Mall, Vancouver, BC, 
V6T 2A3, Canada}
\affiliation{Department of Physics and Astronomy, University of British Columbia, 6224 Agricultural Road, Vancouver, BC, V6T 1Z1, Canada}

\date{\today}

\begin{abstract}

There exist experimental hints from the $B$ sector for CP violation beyond the Standard Model (SM) CKM paradigm. An anomalous dimuon asymmetry was reported by the D0 collaboration, while tension exists between $B \to \tau \nu$ and $S_{\psi K}$.  These measurements, disfavoring the SM at the $\sim 3\sigma$ level, can be explained by new physics in both $B_d^0$- $\bar{B}_d^0$ and $B_s^0$- $\bar{B}_s^0$ mixing, arising from (1) new bosonic degrees of freedom at or near the electroweak scale, and (2) new, large CP-violating phases.   These two new physics ingredients are precisely what is required for electroweak baryogenesis to work in an extension of the SM.  We show that a simple two Higgs doublet model with top-charm flavor violation can explain the $B$ anomalies and the baryon asymmetry of the Universe.  
Moreover, the presence of a large relative phase in the top-charm Yukawa coupling, favored by $B_{d,s}^0$-$\bar{B}_{d,s}^0$ mixing, weakens constraints from $\epsilon_K$ and $b \to s \gamma$, allowing for a light charged Higgs mass of $\mathcal{O}$(100 GeV).

\end{abstract}

\maketitle

\section{Introduction}\label{intro}

Precision tests of CP violation have shown a remarkable consistency with the Standard Model (SM), where all CP-violating observables are governed uniquely by the single phase in the Cabibbo-Kobayashi-Maskawa (CKM) matrix~\cite{Kobayashi:1973fv}.  Yet the search continues.  
Many well-motivated extensions of the SM, such as supersymmetry, contain new sources of CP violation at the electroweak scale.  Furthermore, new CP violation beyond the CKM phase is likely required to explain the origin of the baryon asymmetry of the Universe.  

Recent analyses have suggested that the CKM paradigm may be in trouble.  First, the D0 collaboration has measured the like-sign dimuon asymmetry, arising from CP violation in the mixing and decays of $B^0_{d,s}$ mesons, in excess over SM prediction at the 3.2$\sigma$ level~\cite{Abazov:2010hv}.  Second, there is tension at the $\sim 3\sigma$ level between the branching ratio for $B^+ \to \tau^+ \nu$ and the CP asymmetry $S_{\psi K}$ in $B_d^0 \to J/\psi\, K$~\cite{Deschamps:2008de,Lunghi:2010gv}.  Additionally, CDF and D0 have measured the CP asymmetry $S_{\psi \phi}$ in $B_s^0 \to J/\psi \, \phi$.  While their earlier results (each with $2.8\, \textrm{fb}^{-1}$ data) showed a $\sim 2\sigma$ deviation from the SM~\cite{oldbetas}, this discrepancy has been reduced in their updated analyses with more data ($5.2$ and $6.1 \, \textrm{fb}^{-1}$, respectively)~\cite{newbetas}.  

Although further experimental study is required, taken at face value, these anomalies suggest CP violation from new physics (NP) in the mixing and/or decay amplitudes of $B_d^0$ and $B_s^0$ mesons~\cite{dimuontheoryrefs}.  Recently, the CKMfitter group has performed a global fit to all flavor observables, allowing for arbitrary new physics in $B_{d,s}^0$-$\bar{B}_{d,s}^0$ mixing amplitudes~\cite{Lenz:2010gu}.  They conclude that the SM is disfavored at $3.4\sigma$, while the data seem to favor NP with large CP-violating phases relative to the SM in {\it both} $B_d^0$ and $B_s^0$ mixing.  At the level of effective theory, this NP takes the form 
\be
\mathscr{L}_{\textrm{NP}} \; \sim \; \frac{c_d}{\Lambda^2} \, (\bar b d)^2 + \frac{c_s}{\Lambda^2} \, (\bar b s)^2 \; + \; \textrm{h.c.}  \label{leff}
\ee
These operators can arise from new bosonic degrees of freedom at or near the weak scale, with new large CP-violating phases~\cite{Dobrescu:2010rh,Trott:2010iz,Jung:2010ik,Buras:2010zm}.

It is suggestive that the same NP ingredients, new weak-scale bosons and new CP violation, can also lead to successful electroweak baryogenesis (EWBG).
EWBG, in which the baryon asymmetry is generated during the electroweak phase transition~\cite{Kuzmin:1985mm,Shaposhnikov:1987tw,Cohen:1993nk}, is particularly attractive since two out of three Sakharov conditions~\cite{Sakharov:1967dj} can be tested experimentally.  First, a departure from thermal equilibrium is provided by a strong first-order phase transition, proceeding by bubble nucleation.  While this does not occur in the SM~\cite{SMEWPT}, additional weak-scale bosonic degrees of freedom can induce the required phase transition; these new bosons can be searched for at colliders.  Second, there must exist new CP violation beyond the SM~\cite{SMCPV}.  This CP violation must involve particles with large couplings to the Higgs boson, since it is the interactions of those particles with the dynamical Higgs background field that leads to baryon production.  Precision tests, such as electric dipole moment searches~\cite{Pospelov:2005pr} and flavor observables, can probe directly CP violation relevant for EWBG.
(The third condition, baryon number violation, is provided in the SM by weak sphalerons~\cite{Klinkhamer:1984di}; however, its rate is highly suppressed in all processes of experimental relevance.)

If we wish to connect Eq.~\eqref{leff} to EWBG, it is better to generate these operators at one-loop, rather than tree-level.
Constraints on the mass differences $\Delta M_{d,s}$ in the $B_{d,s}^0$ systems require that $\Lambda^2/|c_{d}| \gtrsim (500\; \tev)^2$ and $\Lambda^2/|c_{s}| \gtrsim (100\; \tev)^2$~\cite{Isidori:2010kg}.  For tree-level exchange, it seems unlikely that all three Sakharov conditions can be met at once.  Sufficient baryon number generation typically requires couplings $\gtrsim \mathcal{O}(10^{-1})$, such that $c_{d,s} \gtrsim \mathcal{O}(10^{-2})$, while a viable phase transition requires $\Lambda \lesssim$ 1 TeV.  Therefore, EWBG requires $\Lambda^2/|c_{d,s}| \lesssim (10 \tev)^2$, at odds with $\Delta M_{d,s}$ constraints.  However, if the operators in Eq.~\eqref{leff} arise at one-loop order, $c_{d,s}$ will have an additional $1/(4\pi)^2$ loop suppression, allowing for both large couplings and lighter scale $\Lambda$, without conflicting with $\Delta M_{d,s}$ constraints.

In this work, we propose that a simple two Higgs doublet model (2HDM) can account for both anomalous CP violation in $B^0_{d,s}$-$\bar{B}^0_{d,s}$ mixing and EWBG.  
Previous works have studied CP violation in $B^0_{d,s}$-$\bar{B}^0_{d,s}$ mixing within a 2HDM~\cite{Dobrescu:2010rh,Trott:2010iz,Buras:2010zm,Jung:2010ik}.  Our setup, described in Sec.~\ref{sec:model}, is different: we assume the NP Higgs doublet \mbox{$(H^+, \, H^0\!+\!iA^0)$} mediates top-charm flavor violation.  In this case, the NP $B^0_{d,s}$-$\bar{B}^0_{d,s}$ mixing amplitudes $(M_{12}^{d,s})_{NP}$ are generated at one-loop order through charge current interactions mediated by $H^+$ (similar to Ref.~\cite{Jung:2010ik}), rather than through tree-level exchange~\cite{Dobrescu:2010rh,Trott:2010iz,Buras:2010zm}.  In Sec.~\ref{sec:flav}, we compute $(M_{12}^{d,s})_{NP}$ in our model.  We find:
\begin{itemize}
\item The best fit values to {\it both} $M_{12}^{d}$ and $M_{12}^{s}$, from Ref.~\cite{Lenz:2010gu}, can be explained in terms of a single NP phase $\vartheta_{tc}$ (defined below).
\item For large values of $\vartheta_{tc}$ prefered by $B^0_{d,s}$-$\bar{B}^0_{d,s}$ mixing observables, constraints from $\epsilon_K$ and $b \to s\gamma$ are weakened and $H^\pm$ can be light ($m_{H^\pm} \sim 100$ GeV).
\end{itemize}

In Sec.~\ref{ewbg}, we discuss in detail EWBG in our 2HDM model.  We focus on the CP violation aspects of EWBG, computing the baryon asymmetry in terms of the underlying parameters of our model by solving a system of coupled Boltzmann equations.  We find that the parameter region favored by flavor observables (specifically, a large $\bar{t}_R t_L H^0$ coupling) can easily account for the observed baryon asymmetry.  However, the relevant CP-violating phase is unrelated to the phase $\vartheta_{tc}$ entering flavor observables.  In Sec.~\ref{conclude}, we summarize our conclusions.


\vspace{2mm}

\section{Model\label{sec:model}}

In a general (type III) two Higgs doublet model~\cite{2HDM}, where both Higgs fields couple to each SM fermion, one can perform a field redefinition such that only one Higgs field acquires a real, positive vacuum expectation value (vev)~\cite{Haber:2006ue}.  We denote the two Higgs doublets by
\be
H_1 = \left( \ba{c} G^+ \\ v + \frac{h^0 + i G^0}{\sqrt{2}} \ea \right) \; , \quad
H_2 = \left( \ba{c} H^+ \\  \frac{H^0 + i A^0}{\sqrt{2}} \ea \right) \; ,
\ee
where $h^0, H^0$ ($A^0$) are the neutral (pseudo)scalars, $H^\pm$ is a charged scalar, and $G^{\pm,0}$ are the Goldstone modes eaten by the electroweak gauge bosons. The vev is $v \approx 174$ GeV.  In general, the physical neutral states can be admixtures of $h^0, H^0, A^0$, depending on the details of Higgs potential.    
We neglect mixing in our analysis; in this case, $H_1$ is exactly a SM Higgs doublet.

The most general Yukawa interaction for $u$-type quarks is
\begin{align}
\mathscr{L}_{\textrm{yuk}} \supset \bar{u}_R (y_{U} H_1 + \widetilde{y}_{U} H_2) Q_L \; +\;  \textrm{h.c.}
\end{align}
where the left-handed quark doublet is $Q_L \equiv (u_L, V d_L)$.  The SU(2)$_L$ contraction is $H_i Q_L \equiv H_i^+ (V d_L) - H_i^0 u_L$.  
The $3\times 3$ Yukawa matrices $y_U$ and $\widetilde y_U$ couple right-handed $u$-type quarks $u_R \equiv (u,c,t)_R$ to left-handed quarks $u_L \equiv (u,c,t)_L$ and $d_L \equiv (d,s, b)_L$.  Working in the mass eigenstate basis, the matrix 
\be
y_U = \textrm{diag}(y_u,y_c,y_t) =\textrm{diag}(m_u,m_c,m_t)/v \label{SMyukawas}
\ee
is a diagonal matrix of SM Yukawa couplings, and $V$ is the CKM matrix.  Analogous Yukawa couplings arise for down quarks and charged leptons:
\begin{align}
&\mathscr{L}_{\textrm{yuk}} \supset \\
&- \bar{d}_R (y_{D} H_1^\dagger + \widetilde{y}_{D} H_2^\dagger) Q_L -\bar{e}_R (y_{L} H_1^\dagger + \widetilde{y}_{L} H_2^\dagger) L_L +  \textrm{h.c.} \notag
\end{align}
where $y_D = \textrm{diag}(y_d,y_s,y_b)$ and $y_L = \textrm{diag}(y_e,y_\mu,y_\tau)$ are the SM Yukawa couplings.

The NP Yukawa matrices $\widetilde y_{U,D,L}$ can be arbitrary.  However, the absence of anomalously large flavor-violating processes provides strong motivation for an organizing principle.  
In this work, we assume that flavor violation arises predominantly in the top sector.  Specifically, we take
\be
\widetilde{y}_U = \left( \ba{ccc} 0 & 0 & 0 \\ 0 & 0 & 0 \\ 0 & \widetilde{y}_{tc} & \widetilde{y}_{tt} \ea \right) \, ,
\quad \widetilde y_{D,L}=0 \label{ansatz} \; .
\ee
That is, we consider a hierarchical structure where the $t_R$-$t_L$ and $t_R$-$c_L$ couplings are dominant (with $|\widetilde y_{tt}| \gg |\widetilde y_{tc}|$), while others are suppressed.  The zeros in Eq.~\eqref{ansatz} are meant to indicate these subleading couplings that for simplicity we neglect in our analysis.   In our setup, flavor violation in meson observables arises at one-loop order through $H^\pm$ charge current interactions, discussed in the next section.

\section{Flavor Constraints\label{sec:flav}}

Mixing and CP violation in the $B_{q}^0$-$\bar{B}_{q}^0$ system ($q\! = \! d,s$) is governed by the off-diagonal matrix element $M_{12}^q - \frac{i}{2} \Gamma^q_{12}$ in the Hamiltonian~\cite{Bigi:2009zz,Buras:1998raa}, with $M_{12}^q$ ($\Gamma_{12}^q$) associated with the (anti-)Hermitian part.  Only the relative phase $\phi_q \equiv \arg(-M_{12}^q/\Gamma_{12}^q)$ is physical.
The relevant observables are the mass and width differences between the two eigenstates
\be
\Delta M_q = 2|M_{12}^q| \, , \quad \Delta \Gamma_q = 2 |\Gamma_{12}^q| \cos\phi_q \; ,
\ee
and the wrong sign semileptonic asymmetry
\be
a_{sl}^q \equiv \frac{\Gamma(\bar{B}_q^0 \to \mu^+ X) - \Gamma(B_q^0 \to \mu^- X)}{\Gamma(\bar{B}_q^0 \to \mu^+ X) + \Gamma(B_q^0 \to \mu^- X)} = \frac{|\Gamma_{12}^q|}{|M_{12}^q|} \, \sin\phi_q \; .
\ee
The dimuon asymmetry measured by D0 arises from wrong sign semileptonic decays of both $B_d^0$ and $B_s^0$ mesons and is given by $A_{sl}^b \approx 0.5 \, a_{sl}^d + 0.5 \, a_{sl}^s$~\cite{Abazov:2010hv}.

In the SM, the mixing amplitude $M_{12}^q$ arises from box graphs, while the $\Gamma_{12}^q$ comes from tree-level decays.  Therefore, it is plausible that NP effects enter predominantly through mixing.  Deviations in $M_{12}^q$ from the SM can be parametrized by 
\be
M_{12}^q = (M_{12}^q)_{SM} + (M_{12}^q)_{NP} \; \equiv \; (M_{12}^q)_{SM} \, \Delta_q \; .
\ee
The consistency of $\Delta M_{d,s}$ with SM predictions constrains $|\Delta_{d,s}| \approx 1$, at the $\mathcal{O}(20\%)$ level~\cite{Lenz:2010gu}, while the dimuon asymmetry measurement disagrees with SM prediction at $3.2\sigma$ and requires $\mathcal{O}(1)$ NP phases $\phi_q^\Delta \equiv \arg(\Delta_q)$~\cite{Abazov:2010hv}.  Phases $\phi_q^\Delta$ also enter into CP asymmetries due to interference between $B^0_{d,s}$ decay amplitudes with and without mixing: e.g., the asymmetry for $B_d^0 \to J/\psi \, K_S^0$ is $S_{\phi K_S} = \sin(2\beta + \phi_d^\Delta)$, with CKM angle $\beta \equiv \arg(-V_{cd}V_{cb}^* V_{td}^* V_{tb})$.  As emphasized in Ref.~\cite{Lunghi:2010gv}, the presence of non-zero $\phi_d^\Delta$ can alleviate tension between $S_{\phi K_S}$ and $\textrm{Br}(B^+ \to \tau^+ \nu)$, which is sensitive to $\beta$ but not $\phi_d^\Delta$.

To quantify these tensions, the CKMfitter group performed a global fit allowing for arbitrary $\Delta_{d,s}$ (dubbed ``Scenario I''), finding that the SM point ($\Delta_d = \Delta_s = 1$) is disfavored at $3.4\sigma$~\cite{Lenz:2010gu}.  Moreover, their best fit point favors NP CP-violating phases in both $B_d^0$ and $B_s^0$ mixing: $\phi_d^\Delta = (-12^{+ 3.3}_{-3.4})^\circ$ and $\phi_s^\Delta = (-129^{+12}_{-12})^\circ \cup (-51.6^{+14.1}_{-9.4})^\circ$.\footnote{Ref.~\cite{Lenz:2010gu} did not include in their fit updated CDF and D0 results for $S_{\phi\psi}$~\cite{newbetas}, which showed improved consistency with the SM over previous results favoring non-zero $\phi_s^\Delta$~\cite{oldbetas}.}

\begin{figure}[t!]
\begin{center}
\includegraphics[scale=.9]{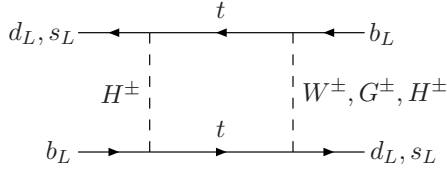} 
\end{center}
\caption{New physics $B_d^0$-$\bar{B}_d^0$ and $B_s^0$-$\bar{B}_s^0$ mixing amplitudes $(M_{12}^{d,s})_{\textrm{NP}}$ arising from box graphs with $H^\pm$ exchange.}
\label{fig:box}
\vspace{-0.5cm}
\end{figure}

In our model, NP effects enter $B_{d,s}^0$ observables predominantly through mixing, via box diagrams shown in Fig.~\ref{fig:box}.  We find\footnote{We neglect running between the scales  $m_t$, $m_W$, and $m_H^\pm$, integrating out these degrees of freedom at a common electroweak scale.  Moreover, we have neglected a NP QCD correction factor $\eta(x_H,x_t)/\eta_B$ arising at next-to-leading order~\cite{Urban:1997gw}.}
\be
\Delta_q = 1 + c_{bq} \, F_1(x_H,x_t)/S_0(x_t) + c_{bq}^2 \, F_2(x_H,x_t)/S_0(x_t) \; , 
\ee
where
\be \label{np}
c_{ij} \equiv \frac{ (\widetilde y_U V)_{ti} (\widetilde y_U V)_{tj}^*}{4\sqrt{2} G_F m_W^2 V_{ti} V_{tj}^*} \; .
\ee
The $\bar{t}_R d_{L}^{\,i} H^+$ charge current couplings are
$(\widetilde y_U V)_{ti} = \widetilde y_{tt} V_{ti} + \widetilde y_{tc} V_{ci}$, for $i=d,s,b$.
The NP loop functions are 
\begin{align}
F_1(x_H,x_t) =& \frac{x_t x_H (x_H-4) \log x_H}{(x_H-1)(x_H-x_t)^2} - \frac{x_t(x_t-4)}{(x_t-1)(x_H-x_t)} \notag \\
              & - \; \frac{x_t(x_H x_t^2 - 2 x_H x_t+4 x_H-3x_t^2) \log x_t }{(x_t-1)^2(x_H-x_t)^2} \\
F_2(x_H,x_t) =& \frac{x_H^2 - x_t^2 - 2 x_t x_H \log(x_H/x_t)}{(x_H-x_t)^3} 
\end{align}
where $x_{t,H} \equiv m_{t,H^\pm}^2/m_W^2$, and $S_0(x_t) \approx 2.35$ is the SM loop function (e.g., see~\cite{Buras:1998raa}).  

$B^0_{d,s}$-$\bar{B}^0_{d,s}$ mixing from box graphs in a 2HDM have been computed previously~\cite{Hou:1988gv}.  Here, a novel feature arises from the NP CP-violating phase associated with $\widetilde{y}_{tc}$~\cite{Diaz:2005rv}.  We can write $(\widetilde y_U V)_{ti}$ as
\begin{align}
(\widetilde y_U V)_{tb} &\simeq \widetilde{y}_{tt} V_{tb}  \notag \\
(\widetilde y_U V)_{ts} &= \widetilde{y}_{tt} V_{ts} \left( 1 + \left| \frac{\widetilde y_{tc}V_{cs}}{\widetilde y_{tt}V_{ts}} \right|  e^{i \vartheta_{tc}} \right)   \\
(\widetilde y_U V)_{td} &= \widetilde{y}_{tt} V_{td} \left( 1 + \left| \frac{\widetilde y_{tc} V_{cd}}{\widetilde y_{tt}V_{td}} \right|  e^{i (\vartheta_{tc}+\beta)} \right) \; ,  \notag
\end{align}
where $\vartheta_{tc} \equiv \arg( \widetilde y_{tc} V_{cs} \widetilde y_{tt}^* V_{ts}^*)$.  In the limit $|\widetilde y_{tt}| \gg |\widetilde y_{tc}|$, we neglect the term $\widetilde y_{tc} V_{cb}$ for $i=b$; however, $y_{tc}$ is non-negligible for $i=d,s$ because the $\widetilde{y}_{tt}$ terms are Cabibbo suppressed.  

The NP phase that enters $(M_{12}^s)_{NP}$ is $\vartheta_{tc}$, while for $(M_{12}^d)_{NP}$ it is $(\vartheta_{tc}\!+\beta)$, due to the different CKM structures of $(\widetilde y_U V)_{ts}$ and $(\widetilde y_U V)_{td}$.  The best fit values for $\phi_{d,s}^\Delta$ are quite different numerically, but due to this extra $e^{i \beta}$, we can explain both $\phi_{d,s}^\Delta$ in terms of the single NP phase $\vartheta_{tc}$.
(For $\widetilde{y}_{tc} = 0$, our model gives $\phi_{d,s}^\Delta=0$, since $(M_{12}^q)_{NP}$ would have the same complex phase $(V_{tb} V_{tq}^*)^2$ as $(M_{12}^q)_{SM}$.)  

\begin{figure*}[t!]
\begin{center}
\includegraphics[scale=.65]{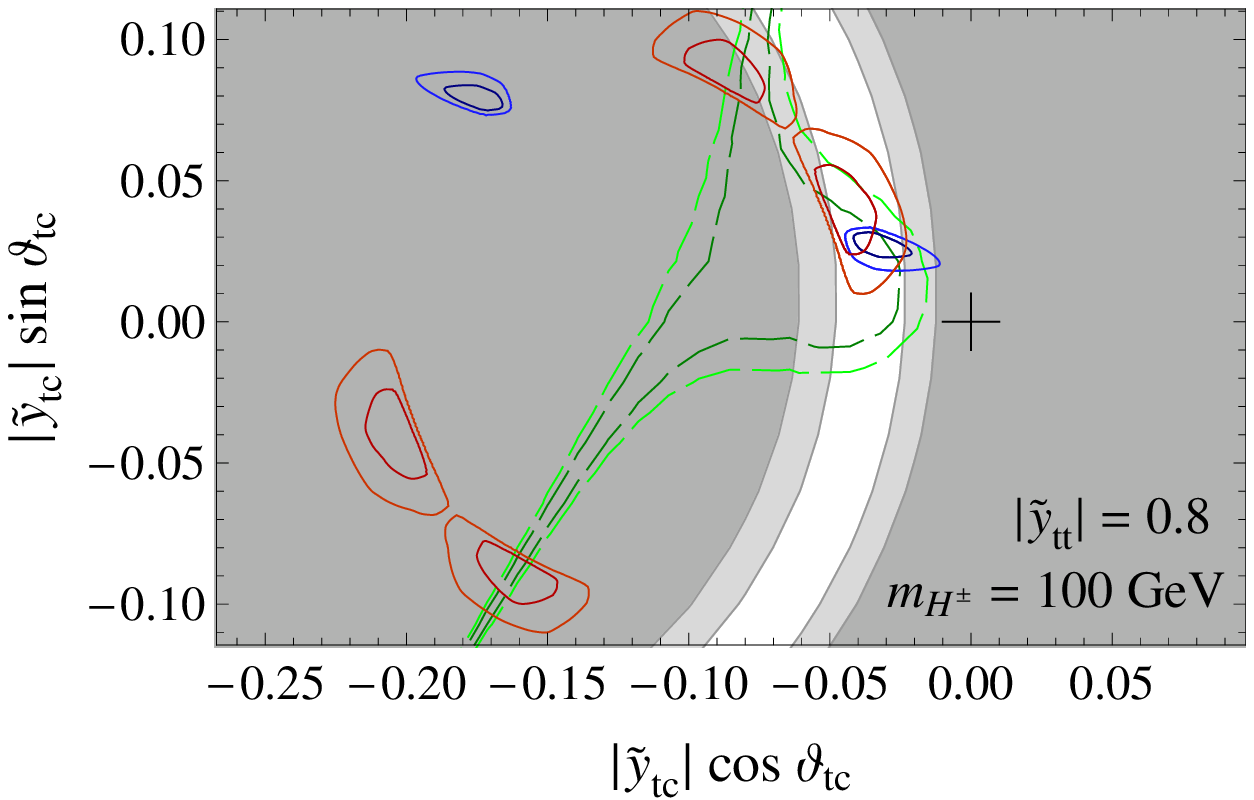}  \quad 
\includegraphics[scale=.65]{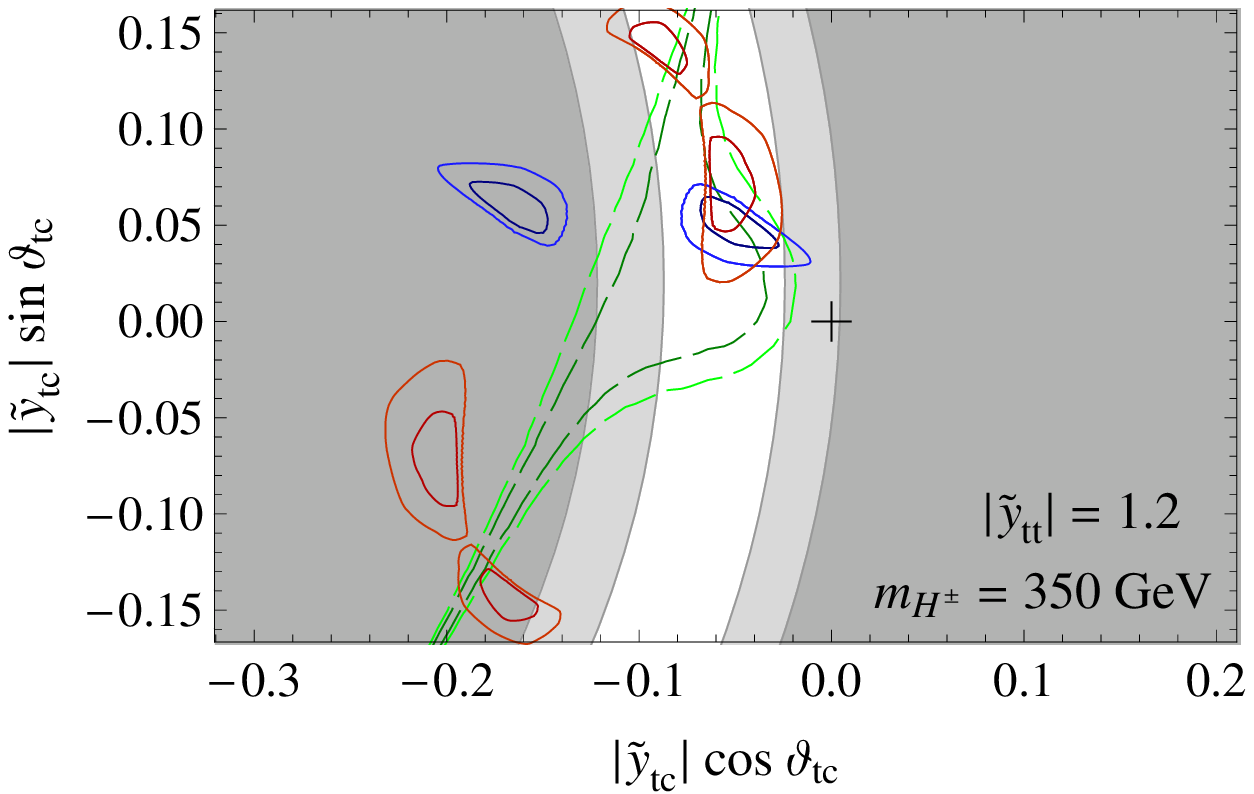} 
\end{center}
\caption{\it \small Top-charm flavor violation parameter space ($|\widetilde y_{tc}|$, $\vartheta_{tc}$) consistent with flavor observables, for two choices of $|\widetilde y_{tt}|$, $m_{H^\pm}$.  68\% and 95\% CL regions for $\Delta_d$ ($\Delta_s$) from Ref.~\cite{Lenz:2010gu} shown by blue (red) contours.  Region within dark (light) dashed green contours is consistent with $\epsilon_K$ at 68\% (95\%) CL.  Light (dark) grey region is excluded at 68\% (95\%) CL from $\textrm{BR}(\bar{B} \to X_s \gamma)$.}
\label{plots}
\vspace{-0.5cm}
\end{figure*}

Our results for $B_{d,s}^0$-$\bar B_{d,s}^0$ mixing are shown in Fig.~\ref{plots}.  Here, we map best fit regions for $\Delta_{d,s}$ from Ref.~\cite{Lenz:2010gu} into the parameter space of our model.  We fix $|\widetilde y_{tt}|$ and $m_{H^\pm}$ and evaluate the prefered regions for $|\widetilde y_{tc}|$ and $\vartheta_{tc}$ consistent with $B_{d,s}^0$-$\bar B_{d,s}^0$ mixing constraints.  (As discussed below, EWBG favors $|\widetilde y_{tt}| \sim 1$ and $m_{H^\pm} \lesssim 500$ GeV.) 
The blue (red) contours correspond to the best fit regions at $1\sigma$ (inner) and $2\sigma$ (outer), for $\Delta_d$ ($\Delta_s$).
Since $\Delta_{d,s}$ are quadratic functions of $|\widetilde y_{tc}| e^{i \vartheta_{tc}}$, the best fit regions for $\Delta_{d,s}$ each map into two best fit regions in $|\widetilde y_{tc}|$, $\vartheta_{tc}$ parameter space.

We also implement constraints on our model from $b \to s \gamma$ and $\epsilon_K$.  The branching ratios for $b \to s \gamma$, as measured experimentally~\cite{Asner:2010qj} and evaluated theoretically in the SM at next-to-leading order (NLO)~\cite{Gambino:2001ew}, are given by\footnote{In the observed value, the first error is experimental, while the second is a theoretical error associated with a photon shape function used to extrapolate the branching ratio to different photon energies $E_\gamma$.  Also, although $\textrm{BR}[\bar{B} \to X_s \gamma]$ has been computed at NNLO in the SM~\cite{Misiak:2006ab}, we work at NLO since 2HDM contributions have been computed at NLO only.}:
\begin{align}
\textrm{BR}[\bar{B} \to X_s \gamma]_{E_\gamma > 1.6 \, \textrm{GeV}}^{\textrm{exp}} &= (3.55 \pm 0.24 \pm 0.09)\times 10^{-4}   \label{bsgam} \\
\textrm{BR}[\bar{B} \to X_s \gamma]_{E_\gamma > 1.6 \, \textrm{GeV}}^{\textrm{SM}} &= (3.60 \pm 0.30)\times 10^{-4} \; .\notag
\end{align}
We evaluate SM+NP contributions to $\textrm{BR}[\bar{B} \to X_s \gamma]$ in our model at NLO following Refs.~\cite{Gambino:2001ew, Ciuchini:1997xe}, except that we take as inputs the best fit CKM parameters given in Table 11 of Ref.~\cite{Lenz:2010gu}.  Adding all errors in Eqs.~\eqref{bsgam} in quadrature, we take the following constraint on our model: 
\be
\textrm{BR}[\bar{B} \to X_s \gamma]_{E_\gamma > 1.6 \, \textrm{GeV}}^{\textrm{SM+NP}} = (3.55 \pm 0.39) \times 10^{-4} \; . \label{bsgamcon}
\ee
In Fig.~\ref{plots}, the white (light grey) region corresponds to $|\widetilde{y}_{tc}|$, $\vartheta_{tc}$ parameter space consistent with Eq.~\eqref{bsgamcon} at less than $1\sigma$ ($2\sigma$), while the dark grey region is excluded at $2\sigma$.

NP contributions to $K^0$-$\bar{K}^0$ mixing arise in our model through box graphs analogous to Fig.~\ref{fig:box}.  The strongest constraint is due to $\epsilon_K$.  In the SM, $|\epsilon_K|_{\textrm{SM}} = (1.90 \pm 0.26)\times 10^{-3}$~\cite{Brod:2010mj}, while experimentally $|\epsilon_K|_{\textrm{exp}} = (2.228 \pm 0.011)\times 10^{-3}$~\cite{Nakamura:2010zzi}.  The SM + NP value of $\epsilon_K$ is
\begin{align}
&|\epsilon_K|_{\textrm{SM}+\textrm{NP}} = \kappa_\epsilon C_\epsilon \widehat{B}_K \, \textrm{Im}\left[(V_{ts} V_{td}^*)^2 \eta_2 \notag\right. \\
&\quad \times (S_0(x_t) + c_{sd} F_1(x_H,x_t) + c_{sd}^2 F_2(x_H,x_t) ) \notag \\
&\quad\;\left. + (V_{cs} V_{cd}^*)^2 \eta_1 S_0(x_c) + 2(V_{cs} V_{cd}^* V_{ts} V_{td}^*) \eta_3 S_0(x_c,x_t) \right] \; ,\label{eq:eps}
\end{align}
where NP enters through the coefficients $c_{sd}$ defined in Eq.~\eqref{np}.  (We neglect NP NLO corrections to $\eta_2$.) The remaining SM input parameters in Eq.~\eqref{eq:eps} are defined and tabulated in Ref.~\cite{Lenz:2010gu}. 
Assuming a theoretical error bar as in Ref.~\cite{Brod:2010mj}, we take the following constraint on our model
\be
|\epsilon_K|_{\textrm{SM}+\textrm{NP}} = (2.23 \pm 0.30) \times 10^{-3} \; . \label{epscon}
\ee
It appears that since $|\epsilon_K|_{\textrm{SM}} < |\epsilon_K|_{\textrm{exp}}$, this constraint would favor a small, positive contribution from NP.  However, $|\epsilon_K|_{\textrm{SM}}$ itself is shifted to a central value $|\epsilon_K|_{\textrm{SM}} = 2.40 \times 10^{-3}$ because the best fit CKM parameters in the presence of NP in $B_{d,s}^0$-$\bar B_{d,s}^0$ mixing (given in Table 11 of Ref.~\cite{Lenz:2010gu}) are different than in a SM-only fit.  As a result, Eq.~\eqref{epscon} favors a small, negative contribution from NP.  In Fig.~\ref{plots}, the parameter region within the dashed dark (light) green contours is consistent with $\epsilon_K$ constraint in Eq.~\eqref{epscon} at $1\sigma$ ($2\sigma$). 

Here, we make several important points.
\begin{itemize}
\item Despite the fact that $\phi_{d}^\Delta$ and $\phi_s^\Delta$ are quite different numerically, there exists regions of parameter space where {\it both} NP in 
$B_d^0$-$\bar B_d^0$ and $B_s^0$-$\bar B_s^0$ can be explained by a single phase $\vartheta_{tc}$.  The $1\sigma$ best fit regions for $\Delta_{d,s}$ overlap within the parameter space of our model (neglecting correlations between $\Delta_d$ and $\Delta_s$).
\item The $\Delta_s$ region that overlaps with the $\Delta_d$ region in Fig.~\ref{plots} corresponds to the $\phi_s^\Delta = (-51.6^{+14.1}_{-9.4})^\circ$ solution.  Therefore, our model predicts $\Delta \Gamma_s > 0$.
\item Although $b \to s \gamma$ and $\epsilon_K$ constrain a large parametric region of our model, these two observables are consistent with observation in regions favored by $B$ mixing observables.
\item A large phase $\vartheta_{tc}$ can weaken $b \to s \gamma$ and $\epsilon_K$ constraints, and a light charged Higgs $(m_{H^\pm} \sim 100$ GeV) is not excluded.
\item The values of $(|\widetilde y_{tt}|, m_{H^\pm})$ shown in Fig.~\ref{plots} are consistent with $R_b \equiv \textrm{BR}[Z \to b\bar{b}]/\textrm{BR}[Z \to \textrm{hadrons}]$ at $95\%$ CL~\cite{Jung:2010ik}.
\end{itemize}
Although we chose only two illustrative values $(|\widetilde y_{tt}|, m_{H^\pm})$ = (0.8, 100 GeV) and (1.2, 350 GeV) in Fig.~\ref{plots}, there exists a consistency region between all these observables for parameters $|\widetilde y_{tt}| \sim 1$, $|\widetilde y_{tc}| \sim 0.05- 0.1$, and $\vartheta_{tc} \sim 3\pi/4$, for $100 < m_{H^\pm} < 500$ GeV.  As we discuss below, EWBG favors $|\widetilde y_{tt}| \sim 1$ and $m_{H^\pm} \lesssim 500$ GeV.

\vspace{2mm}

\section{Electroweak Baryogenesis\label{ewbg}}

Given a NP model, viable EWBG requires: (1) the electroweak phase transition must be strongly first order to prevent washout of baryon number, and (2) CP violation must be sufficient to account for the observed baryon-to-entropy ratio $Y_B^{\textrm{obs}} \approx 9 \times 10^{-11}$.  EWBG in a 2HDM has been studied many times previously~\cite{2hdmewb}.  Most recently, Ref.~\cite{Fromme:2006cm} showed that a strong first order phase transition can occur in a type-II 2HDM for $m_{h^0} \lesssim 200$ GeV and $300 \lesssim m_{H^0} \lesssim 500$ GeV.  Although our 2HDM is not exactly the same as in Ref.~\cite{Fromme:2006cm}, we assume that a strong first order transition does occur.  (The phase transition can also be further strengthed or modified by the presence of scalar gauge singlets~\cite{singlets} or non-renormalizable operators~\cite{higherops}.)

We now study baryon number generation during the phase transition.  The dynamical Higgs fields during the transition gives rise to a spacetime dependent mass matrix $M(x)$ for, e.g., $u$-type quarks:
\be
\mathscr{L}_{\textrm{mass}} = - \bar{u}_R M u_L + \textrm{h.c.} \, , \; M = y_U \, v_{1}(T) + \widetilde{y}_U \, v_{2}(T) 
\ee
where $v_{1,2}(T) \equiv \langle H_{1,2}^0 \rangle_{T\ne 0}$ are the vevs at finite temperature $T \approx 100$ GeV.  At zero temperature, when $v_{1}(T),v_{2}(T) \to v,0$, we recover the usual $T=0$ masses.  However, if $v_{2}(T) \ne 0$, then CP-violating quark charge density can arise from $\widetilde y_U$, as we show below.  Left-handed quark charge, in turn, leads to baryon number production through weak sphalerons.  In previous studies, CP asymmetries were generated by a spacetime-dependent Higgs vev phase, arising from CP violation in the Higgs sector~\cite{2hdmewb, Fromme:2006cm}. Here, we assume that the Higgs potential is CP-conserving, such that $v_{1,2}(T)$ do not have spacetime-dependent phases and can be taken to be real.

Is it plausible that $v_{2}(T) \ne 0$ during the phase transition?  Following~\cite{Trott:2010iz}, the most general potential for $H_{1,2}$ can be written
\begin{align}
V &= \lambda( H_1^\dagger H_1 - v^2)^2 + m_{H_2}^2 H_2^\dagger H_2 + \lambda_1 H_1^\dagger H_1 H_2^\dagger H_2 \notag \\
&+ \lambda_2 H_1^\dagger H_2 H_2^\dagger H_1 + [  \,\lambda_3 (H_1^\dagger H_2)^2 + \lambda_4 H_1^\dagger H_2 H_2^\dagger H_2 \notag \\
&+ \lambda_5 H_2^\dagger H_1 (H_1^\dagger H_1 -v^2) + \textrm{h.c.} ] + \lambda_6 (H_2^\dagger H_2)^2 
\end{align}
Our basis choice that $\langle H_2^0 \rangle_{T\!=\!0} = 0$ requires that no terms linear in $H_2$ survive when $H_1^0 \to v$.  The same statement does not hold at $T \ne 0$ due to thermal corrections to $V$.  First, since we expect $v_{1}(T) \ne v$, terms linear in $H_2$ appear proportional to $\lambda_5$.  Second, top quark loops generate a contribution to the potential $(y_t \widetilde{y}_{tt} T^2  H_1^\dagger H_2/4 + \textrm{h.c.})$, given here in the high $T$ limit, also linear in $H_2$.  A proper treatment of this issue requires a numerical evaluation of the bubble wall solutions of the finite $T$ Higgs potential, which is beyond the scope of this project.  Here, we treat $\tan\beta(T) \equiv v_2(T)/v_1(T)$ as a free parameter\footnote{Although the usual $\tan\beta$ is not physical at $T=0$, the angle $\beta(T)$ between the $T=0$ and $T\ne 0$ vev directions {\it is} physical.}, and we work in the $\beta(T) \ll 1$ limit.  Intuitively, we expect $\beta(T)$ to be suppressed in the limit $m_{H_2}^2 \gg T^2$, since the vev will be confined along the $\langle H_2^0\rangle = 0$ valley.

The charge transport dynamics of EWBG are governed by a system of Boltzmann equations of the form $\dot{n}_a = S^\cpv_a + D_a \nabla^2 n_a  + \sum_b \Gamma_{ab} n_b$~\cite{Cohen:1994ss}.  
Here $n_a$ is the charge density for species $a$.  The CP-violating source $S^\cpv_a$ generates non-zero $n_a$ within the expanding bubble wall, at the boundary between broken and unbroken phases, due to the spacetime-varying vevs $v_{1,2}(T)$.  The diffusion constant $D_a$ describes how $n_a$ is transported ahead of the wall into the unbroken phase, where weak sphalerons are active.  The remaining terms describe inelastic interactions that convert $n_a$ into charge density of other species $b$, with rate $\Gamma_{ab}$.  Our setup of the Boltzmann equations follows standard methods, described in detail in Ref.~\cite{Chung:2009qs}.

Following Ref.~\cite{Cohen:1994ss}, we assume a planar bubble wall geometry, with velocity $v_w \ll 1$ and coordinate $z$ normal to the wall.  The $z\!\! >\! 0$ ($z\!\!<\!0$) region corresponds to the (un)broken phase.  We look for steady state solutions in the rest frame of the wall that only depend on $z$.  Therefore, we replace $\dot{n}_a \to v_w n^\prime_a$ and $\nabla^2 n_a \to n^{\prime\prime}_a$, where prime denotes $\partial/\partial z$.  We adopt kink bubble wall profiles
\begin{align}
v(T)/T &= \xi \,[1+\tanh(z/L_w)]/(2\sqrt{2}) \, , \\
\beta(T) &= \Delta\beta \,[1+\tanh(z/L_w)]/{2}  \;,
\end{align}
where $v(T)^2 \equiv v_1(T)^2 + v_2(T)^2$.
We take $\xi = 1.5$, wall width $L_w = 5/T$, and $T=100$ GeV.  Ref.~\cite{Fromme:2006cm} found viable first-order phase transitions with $1 < \xi < 2.5$ and $2 < L_w T < 15$, depending on the Higgs parameters.  For definiteness, we take $m_{H_2} = 400$ GeV; however, our analysis does not account for the crucially important $m_{H_2}$-dependence of the bubble profiles.

Specializing to our 2HDM, the complete set of Boltzmann equations is
\begin{align}
&v_w n'_{q_a} = D_q n''_{q_a} + \delta_{3a} ( S^\cpv_t  + \Gamma_y Q_y + \Gamma_m Q_m ) - 2 \Gamma_{ss} Q_{ss}\notag\\
&v_w n'_{u_a} = D_q n''_{u_a}  - \delta_{3a} ( S^\cpv_t  + \Gamma_y Q_y + \Gamma_m Q_m ) + \Gamma_{ss} Q_{ss}\notag\\
&v_w n'_{d_a} = D_q n''_{d_a} + \Gamma_{ss} Q_{ss} \label{boltz}\\
&v_w n'_{H} = D_H n''_{H} + \Gamma_y Q_y - \Gamma_h Q_h \notag
\end{align}
with linear combinations of charge densities
\begin{align}
&Q_y \equiv \frac{n_{u_3}}{k_{u_3}} - \frac{n_{q_3}}{k_{q_3}} - \frac{n_{H}}{k_{H}} \, , \quad Q_m \equiv \frac{n_{u_3}}{k_{u_3}} - \frac{n_{q_3}}{k_{q_3}} \\
&Q_{ss} \equiv \sum_{a=1}^3 \left(\frac{2n_{q_a}}{k_{q_a}} - \frac{n_{u_a}}{k_{u_a}} - \frac{n_{u_a}}{k_{u_a}} \right) \, , \quad Q_h = \frac{n_H}{k_H} \; .
\end{align}
The relevant densities are the $a$th generation left(right)-handed quark charges $n_{q_a}$ ($n_{u_a}$, $n_{d_a}$), and the Higgs charge density $n_H \equiv n_{H_1} + n_{H_2}$ (we treat $H_{1,2}$ as mass eigenstates in the unbroken phase).  We assume that (Cabibbo unsuppressed) gauge interactions are in equilibrium, as are Higgs interactions that chemically equilibrate $H_{1,2}$ (provided by $\lambda_{3,4,5}$ quartic couplings in $V$).  Lepton densities do not get sourced and can be neglected.  The $k$-factors are defined by $n_a = T^2 k_a \mu_a/6$, with chemical potential $\mu_a$.

In the Eqs.~\eqref{boltz}, we take these transport coefficients as input:
\begin{align}
&S_t^\cpv \approx 0.1 \times N_c \, |y_t \widetilde y_{tt}| \sin\vartheta_{tt} \,v(T)^2 v_w \beta(T)' \,T \\
&\Gamma_m \approx 0.1 \times N_c |y_t v_1(T) + \widetilde{y}_{tt} v_2(T)|^2  \,T^{-1} \\
&\Gamma_y \approx \frac{27 \zeta_3^2}{2\pi^2} \alpha_s y_t^2 T + 9 |\widetilde{y}_{tt}|^2 T \left(\frac{m_{H_2}}{2\pi T}\right)^{5/2} e^{-m_{H_2}/T} \\
&\Gamma_{ss} \approx 14 \alpha_s^4 T \, , \; D_q \approx 6/T \, , \;  D_H \approx 100/T \; .
\end{align}
We compute the CP-violating source $S^\cpv_t$ and relaxation rate $\Gamma_m$, arising for $t_{L,R}$ only, following the vev-insertion formalism~\cite{Riotto:1998zb,resrelax} (explicit formulae can be found in \cite{Blum:2010by}).\footnote{Although there exist more sophisticated treatments, the reliability of quantitative EWBG computations remains an open question (see discussion in~\cite{Cirigliano:2009yt}).}  The sole source of CP violation here is the phase $\theta_{tt} \equiv \arg(\widetilde y_{tt})$, which is {\it not} the same phase that enters into $B_{d,s}^0$-$\bar B_{d,s}^0$ mixing.\footnote{The reparametrization invariant phase is $\vartheta_{tt} \equiv \arg(\widetilde y_{tt} y_t^* v_1^* v_2)$, but we have adopted a convention where $v_{1,2}(T)$ and $y_t$ are real and positive.}  The dimensionless numerical factors $(0.1)$, obtained following Ref.~\cite{resrelax}, arise from integrals over $t_{L,R}$ quasi-particle momenta, taking as input are the thermal masses (tabulated in~\cite{Chung:2009cb}) and thermal widths ($\gamma_{t_{L,R}} \approx 0.15 g_s^2 T$~\cite{Braaten:1992gd}).  The top Yukawa rate $\Gamma_y$ comes from processes $H_1 t_L \leftrightarrow t_R g$ and $H_2 \leftrightarrow t_R \bar{t}_L$~\cite{Huet:1995sh,Chung:2009cb}.  
The strong sphaleron rate $\Gamma_{ss}$~\cite{Moore:2010jd} plays a crucial role in EWBG in the 2HDM~\cite{Giudice:1993bb}, discussed below, and $D_{q,H}$ are the quark and Higgs diffusion constants~\cite{Joyce:1994zn}.  The relaxation rate $\Gamma_h$ is due to Higgs charge non-conservation when the vev is non-zero.  For simplicity, we set $\Gamma_h = \Gamma_m$~\cite{Cohen:1994ss}; we find deviations from this estimate lead to $\lesssim\mathcal{O}(1)$ variations in our computed $Y_B$.
We have omitted from Eq.~\eqref{boltz} additional Yukawa interactions induced by $y_{tc}$ (e.g., $H_2 \leftrightarrow t_R \bar{c}_L$) because we find they have negligible impact on $Y_B$.  Moreover, CP-violating sources from $y_{tc}$ do not arise at leading order in vev-insertions.  Therefore, $y_{tc}$ plays no role in our EWBG setup (this conclusion may not hold beyond the vev-insertion formalism).

Thus far, we have neglected baryon number violation; this is reasonable since the weak sphaleron rate $\Gamma_{ws} \approx 120 \alpha_w^5 T$~\cite{Bodeker:1999gx} is slow and out of equilibrium.  Therefore, we solve for the total left-handed charge $n_L \equiv \sum_a n_{q_a}$ from Eqs.~\eqref{boltz}, neglecting $\Gamma_{ws}$, and then treat $n_L$ as a source for baryon density $n_B$, according to
\be
v_w n_B' - D_q n_B''  = - (3 \Gamma_{ws} n_L + \mathcal{R} n_B) h\; ,
\ee
with the relaxation rate $\mathcal{R} = (15/4) \Gamma_{ws}$~\cite{Cline:2000nw}.  The sphaleron profile $h(z)$ governs how $\Gamma_{ws}$ turns off in the broken phase~\cite{Dine:1994vf}.  Since the energy of the $T=0$ sphaleron is $E_{\textrm{sph}} \approx 4 M_W/\alpha_w$ , we take~\cite{Braibant:1993is}
\be
h(z) = \exp(-E_{\textrm{sph}}(T)/T) \, , \; \; E_{\textrm{sph}}(T) =  E_{\textrm{sph}} v(T)/v \; .
\ee
Effectively, this cuts off the weak sphaleron rate for relatively small values of the vev: $v(T,z)/T \gtrsim g_2/(8\pi)$.

\begin{figure}[t!]
\begin{center}
\includegraphics[scale=.8]{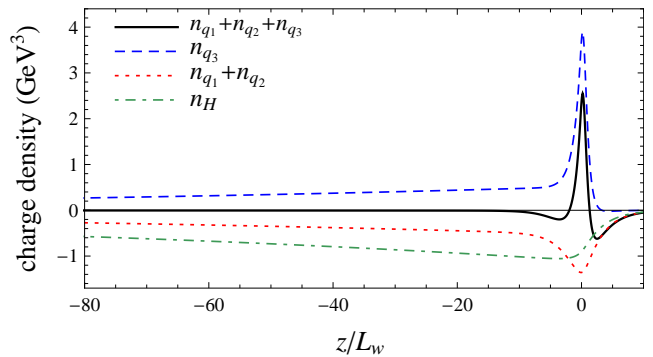}
\end{center}
\vspace{-0.5cm}
\caption{{\it Charge densities in the unbroken phase $(z\!<\!0)$ for $|\widetilde y_{tt}|=1$, $\theta_{tt} = 0.18$, $v_w = 0.05$, $\Delta\beta=10^{-2}$, giving $Y_B \approx 9\times 10^{-11}$.}}
\label{fig:charges}
\end{figure}

In Fig.~\ref{fig:charges}, we show the spatial charge densities resulting from a numerical solution to Eqs.~\eqref{boltz} for an example choice of parameters giving $Y_B \approx 9 \times 10^{-11}$.  In general, the individual charge densities have long diffusion tails into the unbroken phase ($z\!<\!0$).  However, $n_L$ is strongly localized near the bubble wall ($z\!=\!0$), due to strong sphalerons, thereby suppressing $n_B$~\cite{Giudice:1993bb}.  This effect can be understood as follows: at the level of Eqs.~\eqref{boltz}, $B$ is conserved, implying $\sum_a (n_{q_a} + n_{u_a} + n_{d_a}) = 0$; additionally, strong sphalerons relax the linear combination of densities
\be
Q_{ss} \approx (1/N_c) \sum_a (n_{q_a} - n_{u_a} - n_{d_a}) 
\ee
to zero.  These considerations imply that $n_L \approx 0$ if strong sphalerons are in equilibrium.  
In Fig.~\ref{fig:charges}, we see that strong sphalerons are equilibrated and $n_L$ vanishes for $z \lesssim -10L_w$.  
Since $n_L$ is non-zero only near the wall, it is important to treat the weak sphaleron profile accurately in this region, rather than with a simple step function.  Nevertheless, despite this suppression, EWBG can account for $Y_B^{\textrm{obs}}$.  (We also note the significant Higgs charge $n_H$ in the broken phase.  Although we neglect lepton Yukawas here, it is possible that $n_H$ could be efficiently transfered into left-handed lepton charge via $\widetilde{y}_L$, thereby driving EWBG without suffering from strong sphaleron suppression, analogous to Ref.~\cite{Chung:2009cb}.)

\begin{figure}[ttt!]
\begin{center}
\includegraphics[scale=.8]{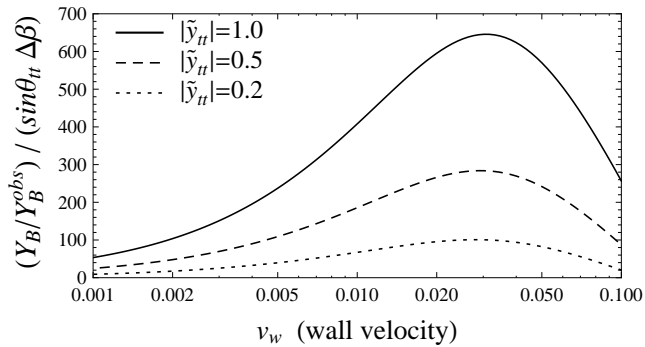}
\end{center}
\vspace{-0.5cm}
\caption{\it Computed baryon asymmetry $Y_B$, normalized with respect to $Y_B^{\textrm{obs}}\sim 9 \times 10^{-11}$ and $(\sin\theta_{tt}\Delta\beta)$, as a function of $v_w$ and $|\widetilde y_{tt}|$.  Vertical axis shows $(\sin\theta_{tt}\Delta\beta )^{-1}$ required for viable EWBG.
}
\label{fog:YB}
\end{figure}

In Fig.~\ref{fog:YB}, we show how large $Y_B$ can be in our model.  The most important parameters are $\Delta\beta$, $\widetilde{y}_{tt}$, and $v_w$ (we find $Y_B$ is not strongly sensitive to $L_w$ or $\xi$).  The vertical axis shows the (inverse) value of $\Delta\beta \times \sin\theta_{tt}$ required for successful EWBG ($Y_B = Y_B^{\textrm{obs}}$), for different values of $|\widetilde{y}_{tt}|$ and $v_w$.  Our main conclusion is that our model can easily account for the baryon asymmetry of the universe -- even if $\Delta\beta$ is as small as $10^{-3} - 10^{-2}$, provided the NP Yukawa coupling has magnitude $|\widetilde{y}_{tt}| \gtrsim 0.2$, with $\mathcal{O}(1)$ phase.  Moreover, $|\widetilde y_{tt}| \sim 1$ is prefered by consistency with flavor observables.

\section{Conclusions\label{conclude}}

The dimuon asymmetry reported by D0~\cite{Abazov:2010hv} and the branching ratio $\textrm{BR}(B \to \tau \nu)$ \cite{Deschamps:2008de,Lunghi:2010gv} seem to disfavor the CKM paradigm of CP violation in the SM at the $\sim 3\sigma$ level.  
Although more experimental scrutiny is required, taken at face value, these anomalies can be accounted for by new physics in both $B_d^0$-$\bar{B}_d^0$ and $B_s^0$-$\bar{B}_s^0$ mixing~\cite{Lenz:2010gu}.  Such new physics would involve new weak-scale bosonic degrees of freedom and new large CP-violating phases.  These two ingredients are precisely what is required for viable electroweak baryogenesis in extensions of the SM.

We proposed a simple 2HDM that can account for these $B$ meson anomalies and the baryon asymmetry.
An interesting feature of our setup is a top-charm flavor-violating Yukawa coupling of the new physics Higgs doublet.  The large relative phase of this coupling can explain {\it both} the dimuon asymmetry and tension in $\textrm{BR}(B \to \tau \nu)$.  Although top-charm flavor violation can give potentially large contributions to $b \to s \gamma$ and $\epsilon_K$ (i.e., less CKM-suppressed than SM contributions), these bounds are weakened in precisely the same region of parameter space consistent with $B_{d,s}^0$-$\bar B_{d,s}^0$ observables.

We also discussed electroweak baryogenesis.  We showed that, provided a strong first-order eletroweak phase transition occurs, our model can easily explain the observed baryon asymmetry of the Universe.  CP violation during the phase transition is provided by the relative phase in the flavor-diagonal $t_L$-$t_R$ Yukawa coupling $\widetilde y_{tt}$ to the new Higgs, and the relevant phase is not related to the top-charm CP phase entering flavor observables.  However, flavor observables and baryogenesis both require $|\widetilde y_{tt}| \sim 1$. Additionally, baryon generation is dependent on a parameter $\Delta \beta$ related to the shift in the ratio of Higgs vevs across the bubble wall.  We expect $\Delta\beta$ to be suppressed in the limit $m_{H^\pm} \gg m_W$. However, we showed that
the charged Higgs state $H^\pm$ can be light $(m_{H^\pm} \sim 100$ GeV) without conflicting with flavor observables due to the large top-charm phase in our model (as opposed to the limit $m_{H^\pm} > 315$ GeV from $b \to s \gamma$ in a type-II 2HDM~\cite{Gambino:2001ew, Nakamura:2010zzi}).  

It would be interesting to explore the consequences of our model for Higgs- and top-related CP-violating and flavor-violating observables measurable in colliders, and also for rare decays such as $K \to \pi \nu \bar \nu$.  Additionally, a more robust analysis of EWBG requires an analysis of the finite temperature effective potential in a Type-III 2HDM, addressing the phase transition strength and bubble wall profiles.

\acknowledgements

We thank the authors of Ref.~\cite{Lenz:2010gu} for sharing with us numerical data from their global fit analysis.
We are also indebted to K. Blum, V. Cirigliano, J. Cline, Y. Hochberg, D. Morrissey, Y. Nir, M. Pospelov, and M. Trott for helpful discussions and suggestions.  S.T. is supported by NSERC of Canada.

\end{document}